\documentclass[11pt]{article}
\usepackage[utf8]{inputenc}
\usepackage[T1]{fontenc}
\usepackage{graphicx}
\usepackage{longtable}
\usepackage{wrapfig}
\usepackage{rotating}
\usepackage[normalem]{ulem}
\usepackage{amsmath}
\usepackage{amssymb}
\usepackage{capt-of}
\usepackage{hyperref}
\usepackage[citestyle=authoryear-icomp,bibstyle=authoryear, hyperref=true,backref=true,maxcitenames=3,url=true,backend=bibtex]{biblatex}
\usepackage{multirow}
\usepackage{amsmath}
\usepackage{morefloats}
\usepackage{setspace}
\usepackage{graphicx}
\usepackage{float}
\usepackage{changepage}
\author{Matt Brigida\thanks{SUNY Polytechnic Institute, 100 Seymour Rd, Utica NY 13502. Email: matthew.brigida@sunypoly.edu}}
\date{\today}
\title{Time-Varying Volatility of Bank Betas}
\hypersetup{
 pdfauthor={Matt Brigida\thanks{SUNY Polytechnic Institute, 100 Seymour Rd, Utica NY 13502. Email: matthew.brigida@sunypoly.edu}},
 pdftitle={Time-Varying Volatility of Bank Betas},
 pdfkeywords={},
 pdfsubject={},
 pdfcreator={Emacs 30.2 (Org mode 9.7.34)}, 
 pdflang={English}}
\begin{document}

\maketitle
\begin{abstract}
Research has shown banks match interest income and expense betas, and thereby obtain net interest income margins which are insensitive to changes in short-term interest rates.  The present analysis extends this research in a number of ways.  First, we use state-space methods to estimate time-varying betas and test whether they are matched at each time interval.  We find substantial variation in interest income and expense betas, which drives variation in net interest margin beta coefficients.  Second, we estimate the time-varying conditional volatility of beta forecasts---the uncertainty of future beta values.  We find uncertainty in interest expense beta coefficients drives uncertainty in interest income betas.  Further, large banks have greater expense beta uncertainty, whereas small banks have greater income beta uncertainty.  Lastly, we find evidence that uncertainty in interest expense betas is priced by the market, and is negatively related to bank stock prices.  This is a new and previously unmeasured source of unhedgeable risk in bank stocks, and highlights an additional benefit of the Federal Reserve's Zero Interest Rate Policy.
\end{abstract}
\vspace*{1cm}


\section{Introduction}
\label{sec:orgac47c89}

Recent research has found evidence that banks hedge interest rate risk by matching the sensitivities of interest income and expense to changes in short-term interest rates (\cite{drechsler2021banking}).  The result is that net interest margins are insensitive to interest rate changes.  Banks are able to do this through a \emph{deposit franchise} which acts like long-term debt rather than the short rate.  That is, the deposit franchise allows banks to maintain interest expense betas which behave similarly to interest income betas.  Therefore, in the presence of a deposit franchise, maturity transformation does not cause interest rate risk.

\cite{drechsler2021banking} estimate static interest income and expense beta coefficients over their entire sample.  There is, however, expected to be substantial variation in interest income betas throughout the period given the sensitivity of duration to the coupon rate and yield.  This raises the question of whether banks are able to match interest income and expense betas over time.  Further, the extent to which the betas are matched does not measure a bank's \emph{uncertainty about whether they will be matched}.  This uncertainty is a yet unmeasured source of bank risk.

Thus matching interest income and expense betas is likely to be done continually. Banks forecast future interest income and expense betas and then adjust their balance sheet to attempt to lessen any difference.  Then banks reforecast betas and adjust their balance sheet in a continual dynamic matching strategy.  A natural model of this process is the Kalman filter, which models a rational market participant which updates forecasts of estimated coefficients in a Bayesian manner as new information arrives in an uncertain environment (\cite{kim2017state}).  

In a similar state-space analysis \cite{brigida2025time} found large banks tend to match income and expense betas at the annual frequency (which may be too infrequently sampled to detect a lead-lag relationship).  Additionally, in their analysis bank betas were calculated using interest income and expense in levels rather than changes. Importantly, they also only considered the point estimate of the deposit beta and ignored the uncertainty in the beta estimate.

Our analysis makes a number of contributions.  First, we show that while banks match interest income and expense betas over time, there is substantial variation in these betas.  Moreover, we find evidence that interest expense betas drive interest income betas.   Second, we estimate the time-varying conditional volatility of beta forecasts, which measures the uncertainty of future beta values.  We find uncertainty in interest expense beta coefficients Granger causes uncertainty in interest income betas.  Lastly, we find evidence that uncertainty in interest expense betas is priced by the market, and is negatively related to bank stock prices.  This is a new and previously unmeasured source of unhedgeable risk in bank stocks.

This uncertainty about beta values peaked during the 2008 financial crisis, and the 2023 regional banking crisis.  There was also substantial uncertainty prior to the 2008 financial crisis, however very low uncertainty during the post-2008 crisis period which was driven by the Federal Reserve's zero interest rate policy.  This latter point shows an additional channel by which the Fed's ZIRP policy was supportive of the banking sector---low beta uncertainty raises bank equity values.  

\cite{drechsler2021banking} found banks can hedge interest rate risk by matching interest income and expense betas.  However our results show that this method of hedging is dynamic, and banks constantly have to match interest income betas to interest expense betas.  This matching process introduces a risk that at a given point the betas will not be matched, and an uncertainty about the ability to match the betas in the future.  Therefore, the ability to hedge interest rate risk is limited through matching betas is limited, and more of an active process than previous research made it seem.

Our analysis contributes to knowledge on the relationship between interest rate changes and bank profitability.  \cite{FLANNERY_1981} and \cite{FLANNERY_1983} find bank profits have little exposure to interest rate changes, however using a data set spanning 10 countries \cite{English2002} finds interest rate changes have a mixed effect on bank profitability.  \cite{Purnanandam_2007} highlights the use of interest rate derivatives to decouple bank lending policy from interest rate shocks.

A separate set of research has focused on the effect of interest rate changes on bank equity.  \cite{English_2018} find evidence that shocks to interest rates do affect bank equity levels, however the effect on banks is only marginally greater than the effect on all firms.  \cite{Begenau_2015} and \cite{Begenau_2018}, however, find evidence that bank balance sheets are significantly exposed to interest rate shocks.

In addition to recent research on matching interest and expense betas (\cite{drechsler2021banking}), bank deposit betas are widely used in academic research on bank market power (\cite{drechsler2017deposits}).  Bank betas are also extensively used in industry and by regulators such as the Federal Reserve Board of Governors (\cite{feds2024}).

Our analysis also contributes to research on the relationship between deposits and policy rate increases. \cite{greenwald2023deposit} find as short-term rates rise, depositors tend to shift to time deposits from savings accounts and similar products.  Moreover, the when bank deposit rates do not increase in kind with short-term interest rates, some depositors switch to money market funds (\cite{xiao2020monetary}). Thus we can expect interest expense betas to rise when interest rates rise.

The remainder of the paper is organized as follows. Section 2 outlines our dataset and empirical methods. Section 3 discusses our results and section 4 concludes.
\section{Data and Methods}
\label{sec:orgdd06d0d}

Our dataset is built from a database of FDIC Call reports and ranges from October 1992 to June 2024.  We collect total interest income to assets (FDIC BankFind code:  \emph{INTINCY}), expense to assets (\emph{EINTEXP}\footnote{This is variable a cumulative annual interest expense, which we difference to obtain the quarterly interest expense.}), and assets for each bank over each quarter.  Federal funds rate data is obtained from the St Louis Federal Reserve Bank's FRED Database.

We analyze bank deciles separately to control for variation in NIM and bank performance driven by bank size.  Nonetheless there is substantial heterogeneity among banks within each decile due to the priorities of each bank's operations.  For example, banks' focusing on credit card operations have a much higher NIM than custody and investment banks.  Regional banks typically have a NIM in between these extremes.  
\subsection{Empirical Methods}
\label{sec:org262d648}

We first estimate constant coefficient regressions, and then test these for parameter instability.  If we can reject constant coefficients, this motivates estimating our interest income and expense betas as time-varying parameters vie state-space methods.  Further, this latter method affords estimates of beta conditional volatility.  The following subsections describe each empirical method.
\subsubsection{Constant Regression Coefficients}
\label{sec:orgc65cb6b}

To calculate the interest income beta we estimate the parameters of the following time-series regression for each decile \emph{d}:

\begin{equation}
\Delta IntInc_{dt} = \alpha^{Inc}_d + \beta^{Inc}_{d,0} \Delta FedFunds_{t} + \beta^{Inc}_{d,1} \Delta FedFunds_{t-1} + \epsilon_{dt}
\end{equation}

and we report:

$$\beta^{Inc}_d = \beta^{Inc}_{d,0} + \beta^{Inc}_{d,1}$$

for each decile \emph{d}.

We calculate interest expense betas in the same fashion.

\begin{equation}
\Delta IntExp_{dt} = \alpha^{Exp}_d + \beta^{Exp}_{d,0} \Delta FedFunds_{t} + \beta^{Exp}_{d,1} \Delta FedFunds_{t-1} + \epsilon_{dt}
\end{equation}

and we report:

$$\beta^{Exp}_d = \beta^{Exp}_{d,0} + \beta^{Exp}_{d,1}$$

for each decile \emph{d}.

Once we have these income and expense beta coefficients for each decile, we can calculate the NIM beta by decile with:

$$\beta^{NIM}_d = \beta^{Inc}_d - \beta^{Exp}_d $$
\subsubsection{Test for Non-Constant Coefficients}
\label{sec:orga080958}

Once we estimate interest expense and interest income beta coefficients, we then test for constant coefficients with the \cite{brown1975techniques} test.  We use the \cite{brown1975techniques} test for a number of reasons.  First, we expect the coefficients to change smoothly through time, rather than discretely (in which case either a \cite{chow1960tests} or \cite{quandt1960tests} test would be appropriate).  Second, given time-varying coefficients are potentially driven by changes in the underlying interest rates, coefficients which change according to a random walk is appropriate.
\subsubsection{Time-Varying Beta Estimates}
\label{sec:orge44002d}

Allowing the parameters of our interest income and expense equations to vary over time affords:

\begin{equation}
\Delta IntInc_{dt} = \alpha_{d,t} + \beta^{Inc}_{d,0, t} \Delta FedFunds_{t} + \beta^{Inc}_{d,1, t} \Delta FedFunds_{t-1} + \epsilon_{dt}
\end{equation}

where coefficients take the form of a random walk:

$$\alpha^{Inc}_{d,t} = \mu_1 + \gamma_1 \alpha^{Inc}_{d, t-1} + \nu_{1,t}$$

$$\beta^{Inc}_{d,0, t} = \mu_2 + \gamma_2 \beta^{Inc}_{d,0, t-1} + \nu_{2,t}$$

$$\beta^{Inc}_{d,1, t} = \mu_3 + \gamma_3 \beta^{Inc}_{d,1, t-1} + \nu_{3,t}$$

$$\epsilon_t \sim i.i.d. N(0, R)$$

$$\nu_t \sim i.i.d. N(0, Q)$$

$$E(\epsilon_t, \nu'_t) = 0$$

and \(\Delta IntInc_{dt}\) is the quarterly change interest income for decile \(d\) at time \(t\) and \(\Delta FedFunds_t\) is the quarterly change in the Federal Funds rate at time \(t\).  \(\beta^{Inc}_{d,0,t}\) is the time-varying coefficient on the contemporaneous Federal Funds rate change for decile \(d\) at time \(t\), and \(\beta^{Inc}_{d,1,t}\) is the coefficient on the Federal Funds rate change lagged one quarter.

The structural form of the time-varying regression coefficients is a random walk.  This form is suggested by \cite{engle1985applications} for cases where market participants modify their estimate of the state solely on the arrival of new information. In addition,  \cite{dangl2012predictive} found that random walk coefficients quickly learn changes in the relationship between model variables.
\subsubsection{Conditional Volatility}
\label{sec:orgef32a57}

In addition to time-varying-parameter coefficients, our model also affords an estimate of conditional volatility through the conditional variance of forecast errors from the Kalman filter (see \cite{Kim_1989}).  Specifically, we estimate the conditional variance as \(H_{t \vert t-1}=x_{t-1}P_{t \vert t-1}x'_{t-1} + \sigma^2_e\) where \(x_{t-1}\) is the vector of the change in the Federal Funds rate and its lag, \(P_{t \vert t-1}\) is the variance-covariance matrix of \(\beta_t\) conditional on information available at time \(t-1\) (\(\beta_{t \vert t-1}\)), and \(\sigma^2_e\) is the variance of the disturbance term.  
\section{Results}
\label{sec:org3aaf54d}

Results are summarized in sections 3.1 through 3.4.  Section 3.1 summarizes the constant coefficient estimates, as well as the tests for time-varying coefficients.  Section 3.2 discusses the results from the state-space formulation with time varying interest expense and income beta coefficients, and also provides results of Granger causality between interest expense and income coefficients.  Section 3.3 provides estimates of time-varying interest income and expense beta conditional volatility, as well as test of Granger causality between interest income and expense volatilities.  Section 3.3 also contains tests of whether this conditional volatility is priced by market participants.
\subsection{Constant Coefficient Regressions}
\label{sec:org56d40d4}

Results from constant coefficient regressions are in table 1 below.  Over our sample period, and every decile, we estimate bank interest expense is slightly more sensitive to the short rate than bank income.  This results in a estimated inverse correlation between the short rate and bank net interest margin.

As bank size increases, the sensitivity of interest expense to the short rate also increases (from 0.1966 in the smallest banks to 0.3423 at the largest).  This is consistent with smaller banks relying on a deposit franchise for funding, while larger banks rely more on capital markets and wholesale deposits which are more sensitive to short rate changes.  Since interest income betas are much more uniform across bank size, NIM betas are lower for larger banks. 

\begin{table}[htbp]
\caption{Interest income, expense and NIM beta estimated from constant coefficient regressions.  Data are quarterly and range from October 1992 to June 2024.}
\centering
\begin{tabular}{rrrr}
\hline
\hline
Decile & Interest Income Beta & Interest Expense Beta & NIM Beta\\
\hline
1 & 0.09143 & 0.1966 & -0.10517\\
2 & 0.09294 & 0.2212 & -0.12826\\
3 & 0.09576 & 0.2396 & -0.14384\\
4 & 0.10307 & 0.2471 & -0.14403\\
5 & 0.10133 & 0.2587 & -0.15737\\
6 & 0.10187 & 0.2624 & -0.16053\\
7 & 0.10519 & 0.2769 & -0.17171\\
8 & 0.10974 & 0.2796 & -0.16986\\
9 & 0.11214 & 0.2993 & -0.18716\\
10 & 0.12661 & 0.3423 & -0.21569\\
\hline
\hline
\end{tabular}
\end{table}

\noindent \emph{Tests for Time-Varying Coefficients} \\

Applying the \cite{brown1975techniques} we are able to reject the null of constant coefficients over all deciles. This evidence motivates the estimation of time-varying coefficients, which vary according to a random walk.
\subsection{Time-Varying Coefficients}
\label{sec:orgf41c204}

Plots of time-varying beta coefficients be decile are in figures 1, 2, and 3 below. Figures 4, 5, and 6 show histograms of the time-varying beta coefficients by decile. Interest income betas typically range between -0.2 and 0.4, though the range differs by decile.  Augmented Dickey-Fuller tests on all time-varying beta coefficient series reject the null, which is evidence the beta coefficient series do not contain a unit root.

\begin{figure}[htbp]
\centering
\includegraphics[width=.9\linewidth]{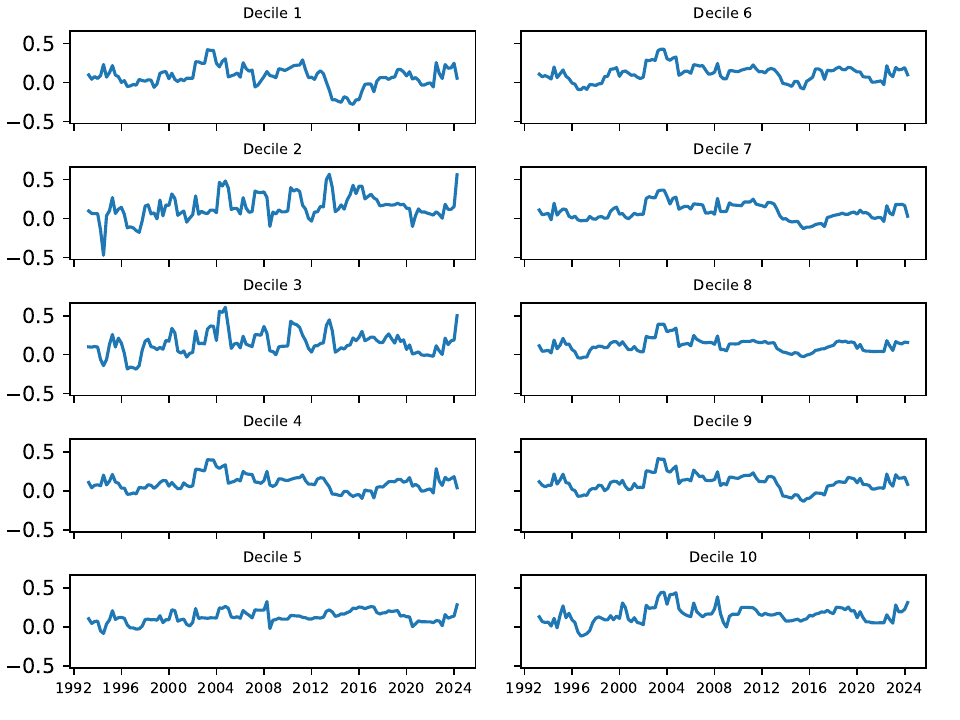}
\caption{Time-Varying Interest Income Beta. The beta was estimated over the sample of quarterly data from October 1992 to June 2024.}
\end{figure}

\begin{figure}[htbp]
\centering
\includegraphics[width=.9\linewidth]{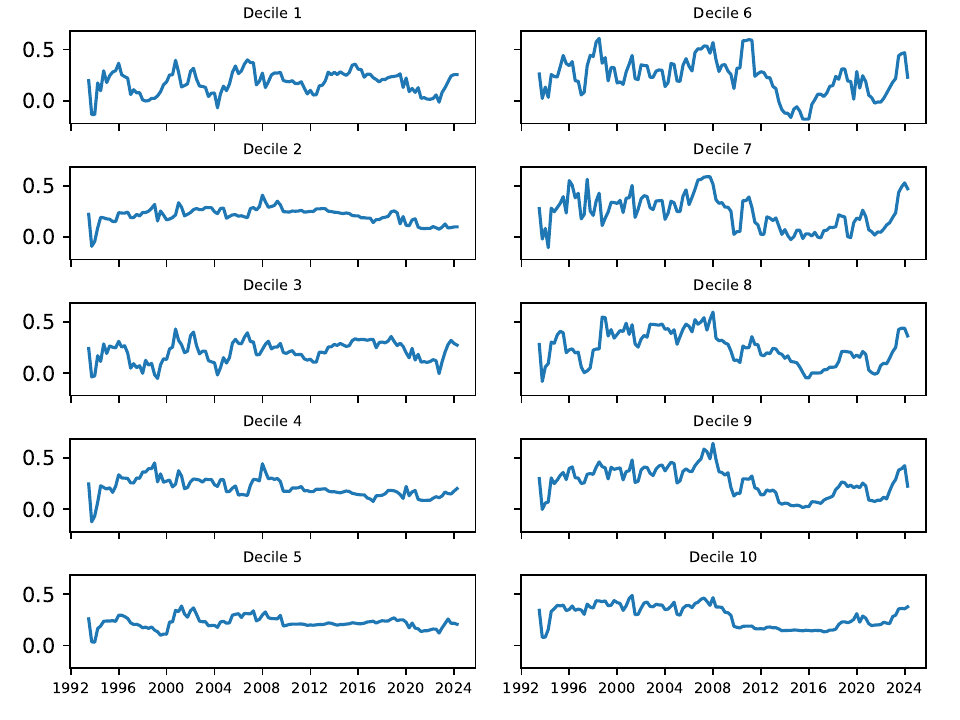}
\caption{Time-Varying Interest Expense Beta. The beta was estimated over the sample of quarterly data from October 1992 to June 2024.}
\end{figure}

\begin{figure}[htbp]
\centering
\includegraphics[width=.9\linewidth]{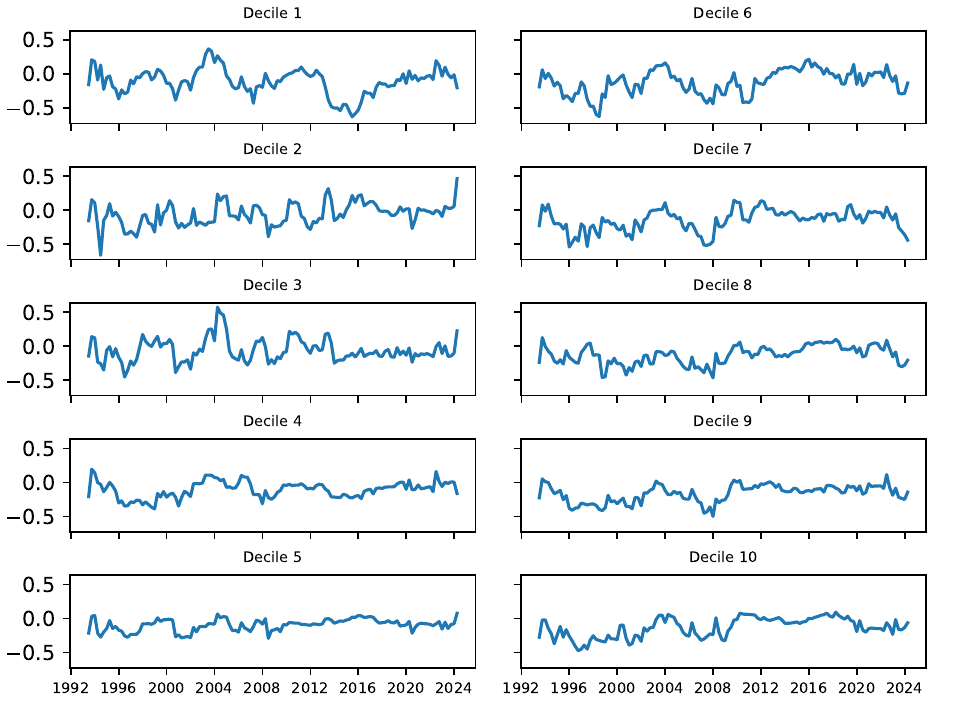}
\caption{Time-Varying Net Interest Margin Beta. The beta was estimated over the sample of quarterly data from October 1992 to June 2024.}
\end{figure}

\begin{figure}[htbp]
\centering
\includegraphics[width=.9\linewidth]{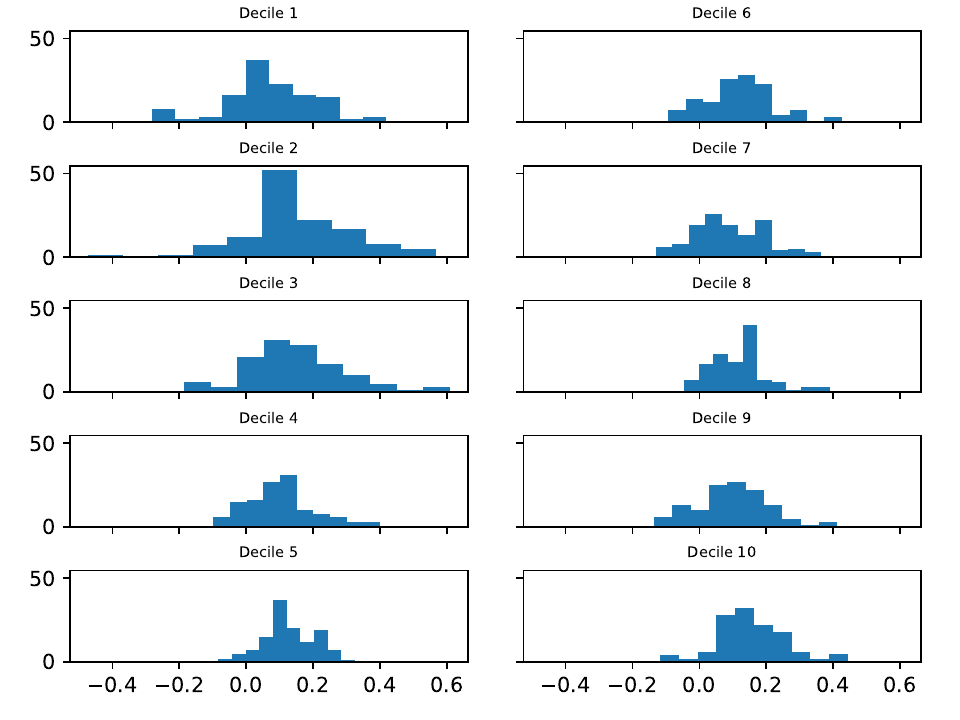}
\caption{Histograms of Time-Varying Interest Income Beta be decile. The beta was estimated over the sample of quarterly data from October 1992 to June 2024.}
\end{figure}

\begin{figure}[htbp]
\centering
\includegraphics[width=.9\linewidth]{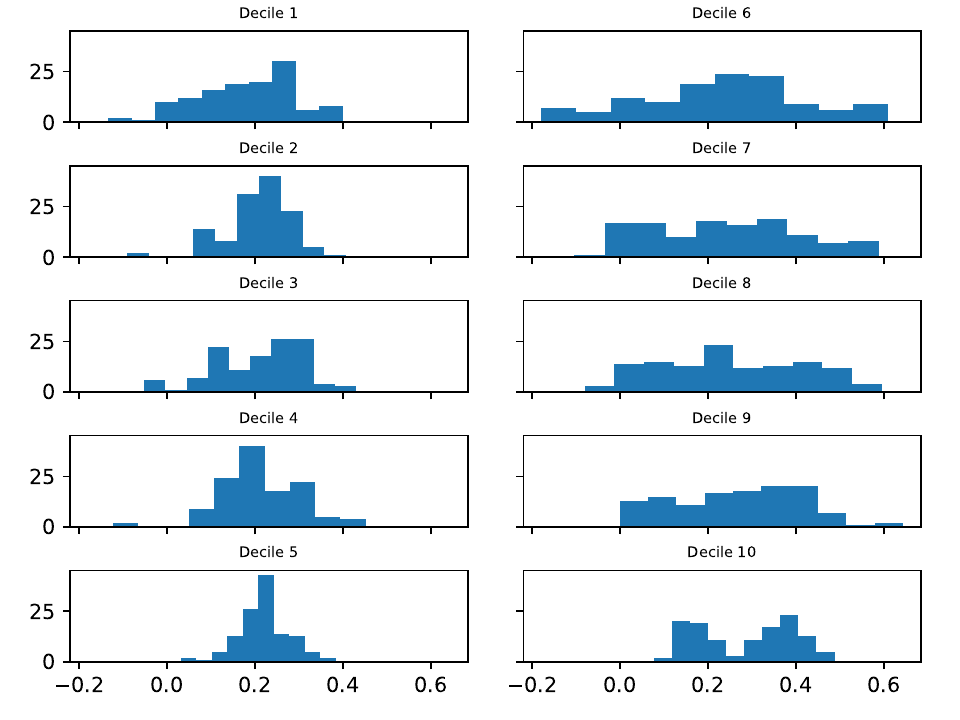}
\caption{Histograms of Time-Varying Interest Expense Beta be decile. The beta was estimated over the sample of quarterly data from October 1992 to June 2024.}
\end{figure}

\begin{figure}[htbp]
\centering
\includegraphics[width=.9\linewidth]{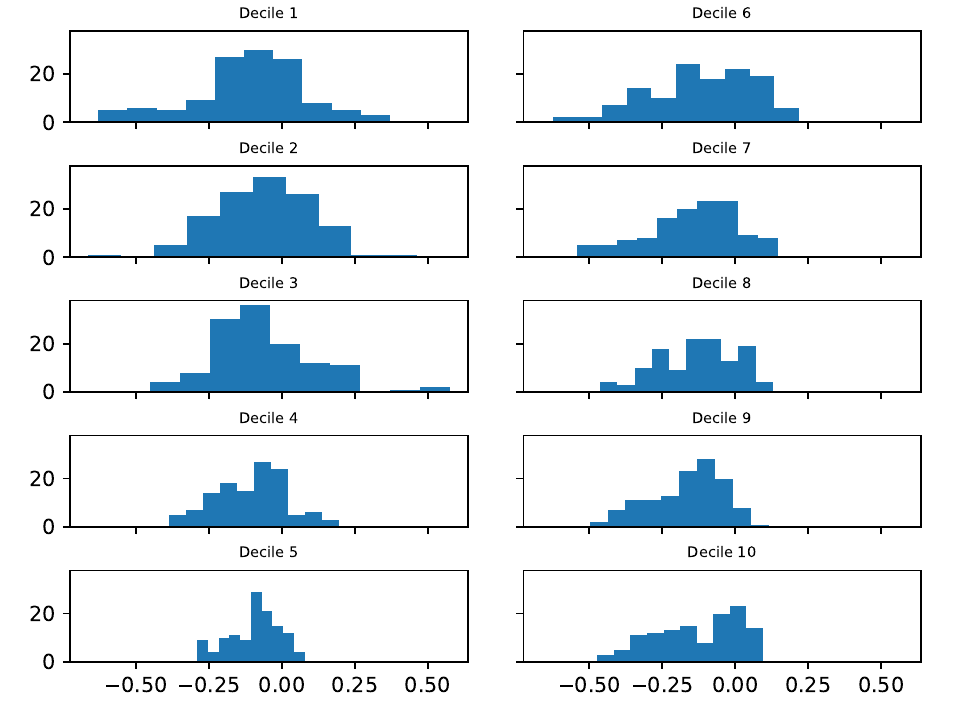}
\caption{Histograms of Time-Varying NIM Beta be decile. The beta was estimated over the sample of quarterly data from October 1992 to June 2024.}
\end{figure}
\subsubsection{Granger Causality}
\label{sec:orge59d9a0}

Table 2 below provides results from Granger causality tests on time-varying interest expense and income beta coefficients. We find some evidence (at the 10\% level of significance) over 4 of 10 deciles that interest expense beta coefficients Granger cause interest income betas.  This is consistent with interest expense sensitivities being driven by exogenous shocks, such as Federal Finds rate changes, and interest income betas being changed to match the expense betas.  There is no evidence of interest income betas Granger causing expense betas.

\begin{table}[htbp]
\caption{Granger Causality tests. Beta coefficient are in levels and are quarterly ranging from October 1992 to June 2024.  Granger causality results are from the SSR based F-test with 4 lagged quarters.}
\centering
\begin{tabular}{rll}
\hline
\hline
Decile & \(\beta_{ii} \Rightarrow \beta_{ie}\) & \(\beta_{ie} \Rightarrow \beta_{ii}\)\\
\hline
1 & 0.3532 & 1.7442\\
 & (0.8413) & (0.1453)\\
\hline
2 & 0.3757 & 1.4759\\
 & (0.8256) & (0.2143)\\
\hline
3 & 0.4705 & 0.1336\\
 & (0.7573) & (0.9697)\\
\hline
4 & 1.0544 & 2.2982\\
 & (0.3827) & (0.0634)\(^*\)\\
\hline
5 & 1.0130 & 0.7629\\
 & (0.4039) & (0.5516)\\
\hline
6 & 0.4154 & 1.1725\\
 & (0.7972) & (0.3269)\\
\hline
7 & 0.4006 & 2.0748\\
 & (0.8078) & (0.0889)\(^*\)\\
\hline
8 & 1.3051 & 2.0697\\
 & (0.2725) & (0.0896)\(^*\)\\
\hline
9 & 0.1652 & 2.1506\\
 & (0.9556) & (0.0793)\(^*\)\\
\hline
10 & 0.1788 & 0.8444\\
 & (0.9489) & (0.4999)\\
\hline
\hline
\end{tabular}
\end{table}
\subsection{Conditional Volatility}
\label{sec:orgaf2baf2}

Plots of the conditional volatility of interest income and expense are in figures 7 and 8 below.  Notably, across deciles, there are peaks in volatility at the 2008 crisis, and the 2023 regional bank crisis.  There is also a smaller increase in volatility around the 2000 technology bubble crash.  Interestingly, volatility was lowest during the post-2008 crisis period, which was dominated by the zero interest rate policy and quantitative easing.

Interest expense beta uncertainty generally peaks a year after interest income beta uncertainty.  This may be because going into a crisis a bank becomes uncertain about how sensitive their income will be to interest rates, whereas coming out of the crisis banks are uncertain how sensitive their expenses will be to interest rates.

\begin{figure}[htbp]
\centering
\includegraphics[width=.9\linewidth]{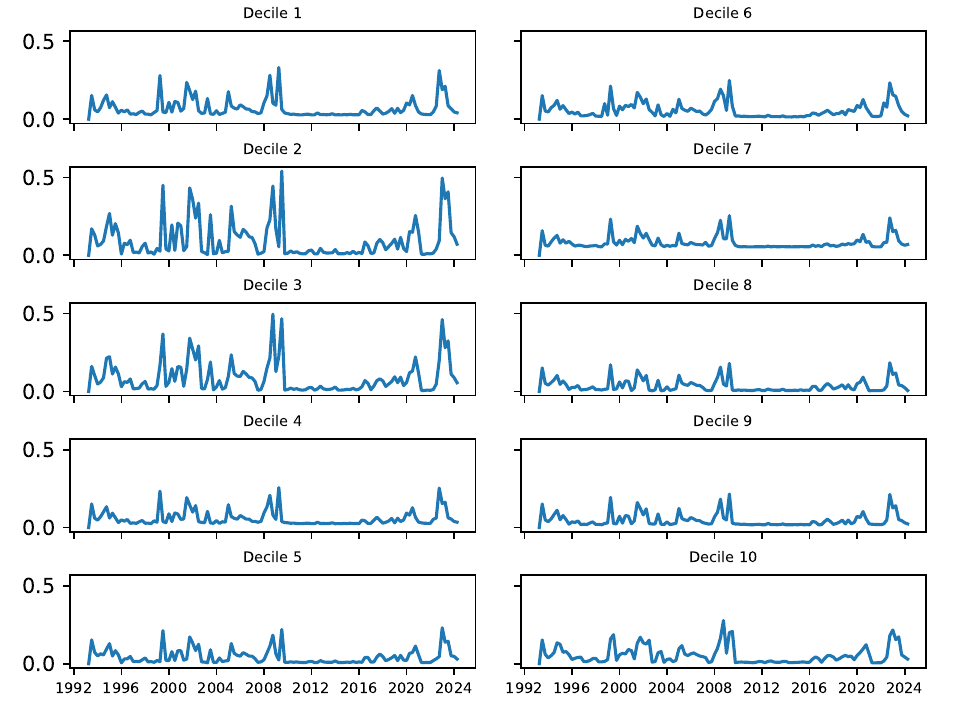}
\caption{Time-Varying Conditional Volatility of the Interest Income Beta. The beta was estimated over the sample of quarterly data from October 1992 to June 2024.}
\end{figure}

\begin{figure}[htbp]
\centering
\includegraphics[width=.9\linewidth]{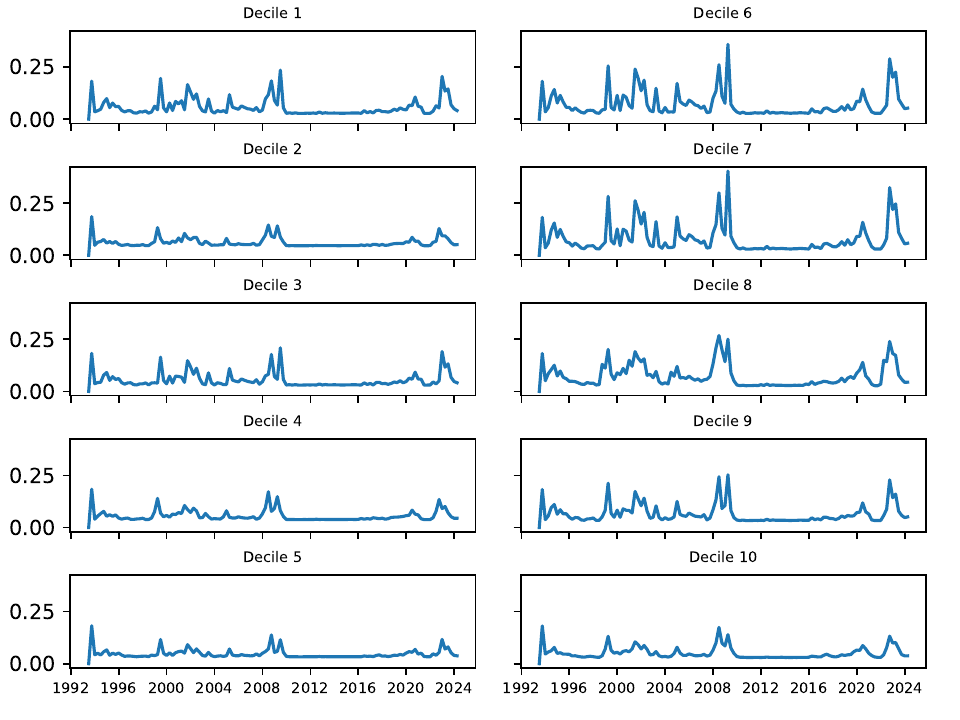}
\caption{Time-Varying Conditional Volatility of the Interest Expense Beta. The beta was estimated over the sample of quarterly data from October 1992 to June 2024.}
\end{figure}

\begin{figure}[htbp]
\centering
\includegraphics[width=.9\linewidth]{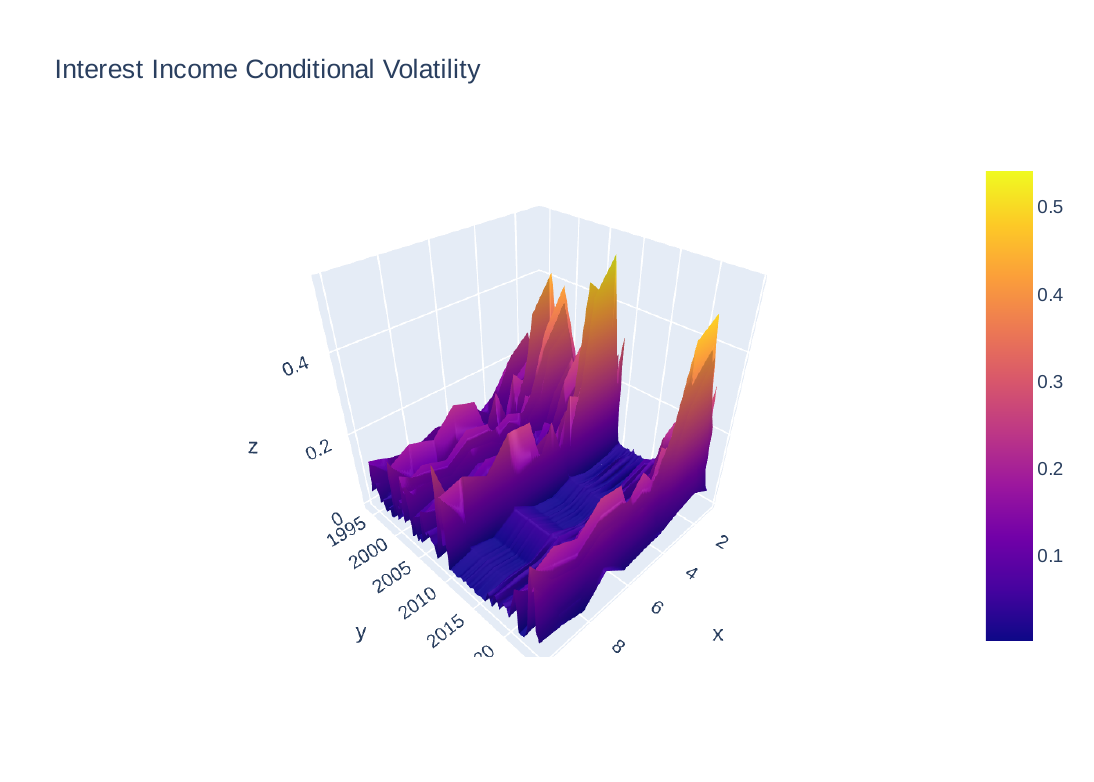}
\caption{Time-Varying Conditional Volatility of the Interest Income Beta. The beta was estimated over the sample of quarterly data from October 1992 to June 2024.}
\end{figure}

\begin{figure}[htbp]
\centering
\includegraphics[width=.9\linewidth]{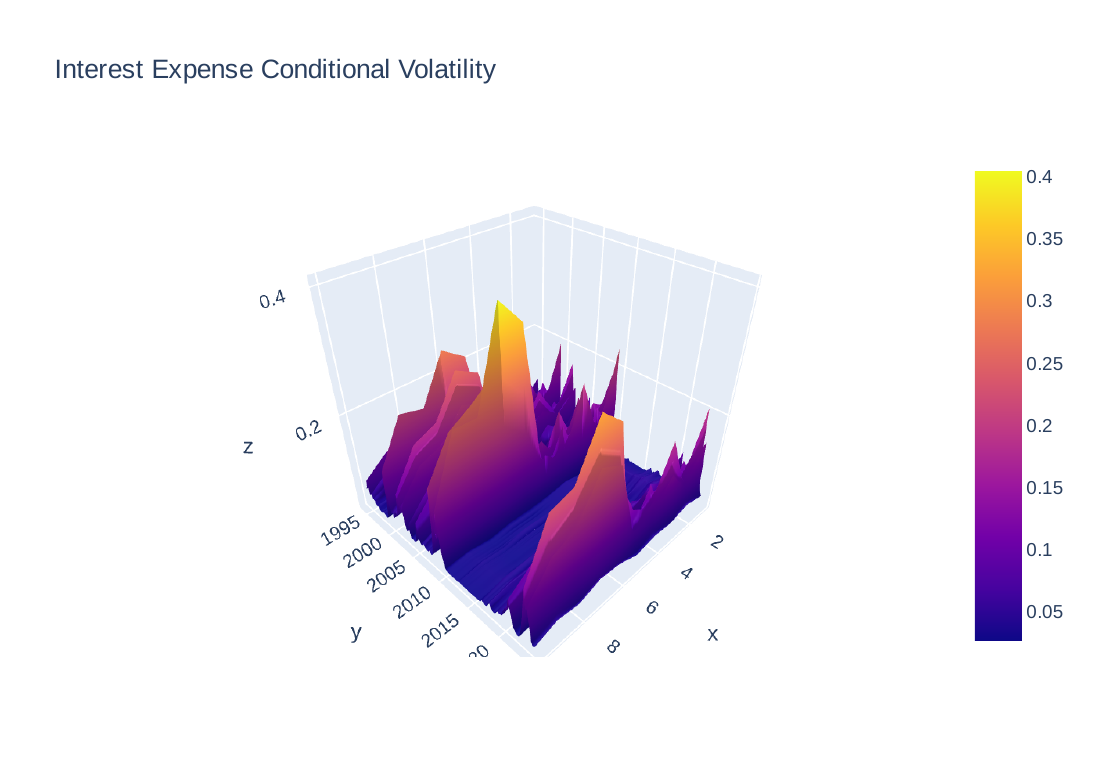}
\caption{Time-Varying Conditional Volatility of the Interest Expense Beta. The beta was estimated over the sample of quarterly data from October 1992 to June 2024.}
\end{figure}
\subsubsection{Large vs Small Bank Conditional Volatility}
\label{sec:org50ce661}

Figure 9 and 10 show an interesting relationship between bank size and interest income and expense conditional volatility.  Figure 9 shows smaller banks have much higher interest income beta uncertainty relative large banks.  Conversely, figure 10 shows larger banks have much higher levels of interest expense beta uncertainty relative to small banks.

These figures also show notable peaks prior to the 2008 financial crisis and the 2023 regional banking crisis.  Also, volatility was generally higher prior to the 2008 financial crisis, however was exceptionally low and stable in the aftermath of the crisis.  This may be due to the Federal Reserves zero interest rate policy.  Descriptive statistics for interest income and expense beta volatility are in tables 3 and 4 below.

These results are consistent with smaller banks relying on a strong deposit franchise to keep expenses low.  That is, the profit strategy of small banks is to control interest expense.  This is sensible given their assets are generally mortgage loans and banks have little pricing power on mortgage rates.  Large banks, however, rely less on low funding rates and more on income generated through credit card, investment banking, and other higher margin sources.

\begin{table}[htbp]
\caption{Descriptive Statistics: Interest income beta conditional volatility by decile.  There are 125 quarters.}
\centering
\begin{tabular}{lrrrrrrrrrr}
\hline
 & 1 & 2 & 3 & 4 & 5 & 6 & 7 & 8 & 9 & 10\\
\hline
mean & 0.0700 & 0.0958 & 0.0882 & 0.0585 & 0.0443 & 0.0570 & 0.0808 & 0.0352 & 0.0464 & 0.0546\\
std & 0.0605 & 0.1163 & 0.1004 & 0.0480 & 0.0479 & 0.0483 & 0.0399 & 0.0397 & 0.0418 & 0.0555\\
min & 0.0000 & 0.0000 & 0.0000 & 0.0000 & 0.0000 & 0.0000 & 0.0000 & 0.0000 & 0.0000 & 0.0000\\
25\% & 0.0322 & 0.0152 & 0.0165 & 0.0280 & 0.0111 & 0.0212 & 0.0560 & 0.0080 & 0.0207 & 0.0123\\
50\% & 0.0468 & 0.0445 & 0.0574 & 0.0395 & 0.0235 & 0.0421 & 0.0662 & 0.0169 & 0.0278 & 0.0375\\
75\% & 0.0800 & 0.1312 & 0.1147 & 0.0629 & 0.0578 & 0.0741 & 0.0872 & 0.0470 & 0.0522 & 0.0694\\
max & 0.3316 & 0.5407 & 0.4960 & 0.2568 & 0.2305 & 0.2489 & 0.2548 & 0.1838 & 0.2153 & 0.2773\\
\hline
\end{tabular}
\end{table}

\begin{table}[htbp]
\caption{Descriptive Statistics: Interest expense beta conditional volatility by decile.  There are 125 quarters.}
\centering
\begin{tabular}{lrrrrrrrrrr}
\hline
 & 1 & 2 & 3 & 4 & 5 & 6 & 7 & 8 & 9 & 10\\
\hline
mean & 0.0566 & 0.0607 & 0.0539 & 0.0559 & 0.0465 & 0.0702 & 0.0775 & 0.0736 & 0.0649 & 0.0507\\
std & 0.0408 & 0.0233 & 0.0355 & 0.0266 & 0.0222 & 0.0609 & 0.0676 & 0.0537 & 0.0444 & 0.0286\\
min & 0.0000 & 0.0000 & 0.0000 & 0.0000 & 0.0000 & 0.0000 & 0.0000 & 0.0000 & 0.0000 & 0.0000\\
25\% & 0.0304 & 0.0467 & 0.0324 & 0.0396 & 0.0342 & 0.0326 & 0.0347 & 0.0346 & 0.0366 & 0.0323\\
50\% & 0.0415 & 0.0518 & 0.0416 & 0.0459 & 0.0392 & 0.0470 & 0.0532 & 0.0547 & 0.0492 & 0.0401\\
75\% & 0.0621 & 0.0655 & 0.0583 & 0.0619 & 0.0509 & 0.0787 & 0.0906 & 0.0886 & 0.0742 & 0.0583\\
max & 0.2354 & 0.1857 & 0.2093 & 0.1840 & 0.1820 & 0.3590 & 0.4040 & 0.2683 & 0.2531 & 0.1812\\
\hline
\end{tabular}
\end{table}
\subsubsection{Tests of Conditional Volatility Granger Causality}
\label{sec:org33982d3}

Table 3 reports Granger Causality results between interest income and expense beta conditional volatilities.  We find evidence of bi-directional (or mutual) Granger causality between interest expense beta conditional volatility and that of interest income.  Over every decile interest expense beta volatility Granger causes interest income, and over all but 3 deciles interest income Granger causes interest expense.  The interpretation of these results is that either series will help predict the other series, however this evidence is consistent with uncertainty in one variable causing uncertainty in the other as banks try and match these two betas.

\begin{table}[htbp]
\caption{Granger Causality tests. The volatility series are in levels and are quarterly ranging from October 1992 to June 2024.  Granger causality results are from the SSR based F-test with 4 lagged quarters.}
\centering
\begin{tabular}{rll}
\hline
\hline
Decile & \(\sqrt{H_{ii}} \Rightarrow \sqrt{H_{ie}}\) & \(\sqrt{H_{ie}} \Rightarrow \sqrt{H_{ii}}\)\\
\hline
1 & 495.5744 & 8.1844\\
 & (0.0000)\(^{****}\) & (0.0000)\(^{****}\)\\
\hline
2 & 0.9632 & 61.1280\\
 & (0.4307) & (0.0000)\(^{****}\)\\
\hline
3 & 30.3265 & 15.5039\\
 & (0.0000)\(^{****}\) & (0.0000)\(^{****}\)\\
\hline
4 & 1.7093 & 6.2638\\
 & (0.1530) & (0.0001)\(^{****}\)\\
\hline
5 & 0.8505 & 2.3117\\
 & (0.4962) & (0.0621)\(^{*}\)\\
\hline
6 & 58.5093 & 40.0344\\
 & (0.0000)\(^{****}\) & (0.0000)\(^{****}\)\\
\hline
7 & 3.5682 & 2.7466\\
 & (0.0089)\(^{***}\) & (0.0319)\(^{**}\)\\
\hline
8 & 9.8772 & 18.9940\\
 & (0.0000)\(^{****}\) & (0.0000)\(^{****}\)\\
\hline
9 & 2.4967 & 4.1224\\
 & (0.0468)\(^{**}\) & (0.0038)\(^{***}\)\\
\hline
10 & 5.0288 & 97.6406\\
 & (0.0009)\(^{****}\) & (0.0000)\(^{****}\)\\
\hline
\hline
\end{tabular}
\end{table}
\subsubsection{The Market Price of Beta Uncertainty}
\label{sec:org860d277}

A natural question is whether market participants incorporate interest income and expense beta uncertainty into bank stock prices.  If so, we should expect stock prices to decline when uncertainty rises. To begin to answer this question we estimate the following regression:

\begin{equation}
r_{XLF, t} = \gamma_0 + \gamma_1 \Delta CV_{Exp, t} + \gamma_2 \Delta CV_{Inc, t} + \gamma_3 r_{M, t} + \xi_t
\end{equation}

where \(XLF\) are returns on the market-capitalization-weighted Financial Select Sector SPDR Fund over quarter \(t\). \(CV_{Exp,t}\) and \(CV_{Inc,t}\) are quarterly interest expense and income beta conditional volatility for decile 10 over quarter \(t\).  We use decile 10 because these banks dominate the market capitalization of \(XLF\).  \(r_{M,t}\) are quarterly returns on the SPDR S\&P 500 ETF Trust (ticker \(SPY\)).  The parameters of the regression are estimated from the first quarter of 1999 to the first quarter of 2024.

\begin{table}[htbp]
\caption{Estimated coefficients from estimating equation 4 above via OLS.  Data are quarterly and range from Q1 1999 to Q1 2024.}
\centering
\begin{tabular}{lrl}
\hline
\hline
Parameter & Coefficient & value\\
\hline
\(\gamma_0\) & -0.0063 & 0.3191\\
\(\gamma_1\) & -1.1454 & 0.0006****\\
\(\gamma_2\) & 0.0926 & 0.4705\\
\(\gamma_3\) & 1.1078 & 0.0000****\\
\hline
\(Adj.\ R^2\) & 0.7206 & \\
\hline
\hline
\end{tabular}
\end{table}

The coefficient on interest expense beta uncertainty is negative at significant at the 0.1\% level.  This is evidence that uncertainty regarding how interest expense will react to changes in the short rate is indeed priced by market participants.  Further, the negative sign indicates as this uncertainty increases, bank stock returns decline.  The standard deviation of the change in interest expense beta conditional volatility is 0.0205, and given our estimated coefficient on CV of -1.1454, this means if there is a one standard deviation increase in conditional beta volatility large bank market capitalizations will decline by 2.34\%. Given the 10 largest banks have a collevtive market capitalization of approximately \$2 trillion, a one standard deviation increase in interest expense beta uncertainty lowers the largest bank stocks by about \$47 billion.

Interestingly, the coefficient on interest income uncertainty is insignificant.  This is evidence that market participants react to uncertainty in how bank expenses will react to the short rate, though not how income will react to the short rate.  Given we find greater evidence that interest expense beta uncertainty Granger cause interest income beta uncertainty than vice versa, it may be that the market is reacting to the first increase in uncertainty. 
\section{Conclusions}
\label{sec:orgc3120ba}

It was commonly assumed that by borrowing short-term and lending long-term, thereby creating a duration mismatch between their assets and liabilities, banks exposed themselves to substantial interest rate risk.  Recent research, however, has shown banks match the sensitivities of their interest income and expense to the short rate, and so doing obtain net interest income margins which are generally insensitive to changes in short-term interest rates.

This matching of interest income and expense sensitivities (betas) is more likely to be a dynamic interest rate risk strategy, rather than static.  As interest rate levels change, the duration of assets changes and so do the sensitivities of interest income to the short rate.  Therefore, match must constantly adjust their assets and liabilities to match their sensitivities in the presence of a stochastic short rate.  Consistent with this hypothesis, we find evidence for interest income and expense betas which vary according to a random walk, and weak evidence that interest expense betas may Granger cause interest income betas.

We also estimate the conditional volatility of our beta forecasts, which provides a number of interesting results.  First, beta uncertainty increased markedly prior to the 2008 financial crisis and 2023 regional banking crisis.  Also, from approximately 2009 to 2019 beta uncertainty became very low across deciles.  This result highlights the role of the Federal Reserve's zero interest rate policy in providing both low and predictable funding rates for banks.  Interestingly, prior to each financial crisis, interest expense beta uncertainty rose most for large banks however interest income beta uncertainty rose most for small banks. This highlights the small bank reliance on their deposit franchise and the large bank focus on generating income.  Consistent with results on the time-varying betas themselves, we find interest expense beta uncertainty Granger causes uncertainty in interest income betas.  This result is consistent with shocks to interest expense betas being transferred to interest income betas.

Lastly, we find evidence that interest expense beta forecast uncertainty is priced by market participants.  In particular, a one standard deviation increase in interest expense beta forecast uncertainty will reduce large bank stock values by 2.34\%.  This is a previously undocumented and unmeasured source of bank risk, and this risk is not hedgeable with current methods and instruments. Given the Federal Reserve's zero interest rate policy significantly reduced interest expense beta uncertainty, we also document an additional avenue of post-2008-crisis support for large banks.

\clearpage

\printbibliography
\end{document}